\documentstyle[preprint,prl,aps]{revtex} 

\begin{document}

\title{MESOSCOPIC QUANTUM CIRCUIT THEORY TO PERSISTENT
CURRENT AND COULOMB BLOCKADE
\footnote{talk at the 
{\bf Fifth International Wigner Symposium, 
Vienna 25-29 August, 1997}}
}

\author{YOU-QUAN LI
\footnote{on leave from the absence 
of Zhejiang University, Hangzhou 310027 China} }

\address{Institute of Theoretical Physics, 
EPFL, CH-1015 Lausanne, Switzerland}

\maketitle
\begin{abstract}

The quantum theory for mesoscopic electric circuit is briefly described. 
The uncertainty relation for electric charge and current 
modifies the tranditional Heisenberg uncertainty relation.
The mesoscopic ring is regarded as a pure L-design, 
and the persistent current is obtained explicitly.
The Coulomb blockade phenomenon appears when applying to the 
pure C-design.
\end{abstract}
\newpage

\section*{}
Owning to the dramatic achievement in nanotechnology,
there have been many studies on mesoscopic physics \cite{Buot}.
In present talk I briefly demonstrate a
quantum mechanical theory for mesoscopic electric 
circuits based on the fact that electronic charge
takes discrete values \cite{LiCh}. As the application
of this approach, 
the persistent current on a mesoscopic ring and the
Coulomb blockade phenomena are formulated from a new
point of view, about which some details are presented. 
Most importantly, it is a physical realization of 
the deformation of quantum mechanics
studied considerably by other authors in mathematical physics.

\section{Quantized Circuit with Charge Discreteness}

In order to taken into account the discreteness of 
electronic charge. we must impose that
the eigenvalues of the self-adjoint operator
$\hat{q}$ (electric charge ) take discrete
values \cite{LiCh}, i.e.
$\hat{q} | q > = n q_e | q >$
($n \in {\sf \, Z \!\!\! Z \, }$, 
$q_e  = 1.602 \times 10^{-19}$ coulomb).
Since the spectrum of charge is discrete, 
the inner product in charge representation 
will be a sum instead of the usual integral and 
the electric current operator $\hat{P}$ will
be defined by the discrete derivatives \cite{Li}
$\nabla_{q_e}$, $\overline{\nabla}_{q_e}$. 
Thus for the mesoscopic quantum electric circuit 
one will have a finite-difference Schr\"{o}dinger 
equation \cite{LiCh}. 
The uncertainty relation for electric charge and current 
modifies the tranditional Heisenberg
uncertainty relation, namely, 
\begin{equation}
\Delta\hat{q}\cdot\Delta\hat{P} \geq \frac{\hbar}{ 2 }
( 1 + \frac{q^2_e }{\hbar^2} < \hat{H}_0 > ).
\label{eq:uncertainty}
\end{equation}
where 
$\hat{H}_0 = -\frac{\hbar^2}{2}
               \nabla_{q_e}
                \overline{\nabla}_{q_e }
= -\frac{\hbar^2}{2q_e}
   (\nabla_{q_e}-\overline{\nabla}_{q_e})$.
The Hamiltonian of quantum LC-design in the presence of 
exterior magnetic flux reads
\begin{equation}
\hat{H}=-\frac{\hbar^2}{2q_e L}
         (D_{q_e}\, - \overline{D}_{q_e})
        +\frac{1}{ 2C} \hat{q}^2 
         +\varepsilon \hat{q}
\label{eq:Hamiltonian}
\end{equation}
where $L$ and $C$ stand for the inductance and the capacity 
of the circuit respectively, $\varepsilon$ represents the
voltage of an adiabatic aource, and the 
covariant discrete derivatives are defined by
\begin{equation}
D_{q_e} := e^{-\frac{q_e}{\hbar}\phi }
     \frac{ \hat{Q} - e^{ i\frac{ q_e}{\hbar} \phi } }
          { q_e } , \,\,\,\,\,\,
\overline{D}_{q_e} := e^{\frac{q_e}{\hbar}\phi }
\frac{ e^{ -i\frac{ q_e}{\hbar} \phi } - \hat{Q^+}
}{ q_e } ,
\label{eq:eg}
\end{equation}
where 
$ \hat{Q}:=e^{iq_e\hat{p}/\hbar}$ 
is a minimum `shift operator' with the property
$\hat{Q}^{+}|n>=e^{i\alpha_{n+1}}|n+1>$
($\alpha_n 's$ are undetermined phases).
The hamiltonian (\ref{eq:Hamiltonian}) is covariant 
under the gauge transformation,
$\hat{G} D_{q_e} \hat{G}^{-1} = D'_{q_e}, \,\,
  \hat{G} \overline{D}_{q_e} \hat{G}^{-1} = 
     \overline{D}'_{q_e}$ 
where 
$\hat{G}:=e^{-i\beta\frac{\hat{q}}{\hbar}}$ 
and the gauge field $\phi$ transforms as 
$\phi \rightarrow \phi' = \phi - \beta$.
The $\phi$ plays the role of the exterioral magnetic 
flux threading the circuit.

\section{Quantum L-design and Persistent Current}

Now we study the Schr\"{o}dinger equation for
a pure L-design in the presence of magnetic flux,
\begin{equation}
-\frac{\hbar^2 }{2q_e L } (D_{q_e}\, - \overline{D}_{q_e} )
| \psi > = E | \psi >.
\label{eq:gh}
\end{equation}
Because its eigenstates can be simultaneous eigenstates of $\hat{p}$,
eq.(\ref{eq:gh})  is solved by the eigenstate 
$|p>=\sum_{n\in{\sf \,Z\!\!\!Z\,}}
\kappa_n e^{i n q_e p/\hbar}|n>$
($\kappa_n :=\exp(i\sum_{j=1}^n \alpha_j)$).
The energy spectrum is easily calculated as
\begin{equation}
E(p,\phi) = \frac{2\hbar}{q^2_e} \sin^2
\left( \frac{q_e}{2\hbar}(p-\phi)
\right)
\end{equation}
which has oscillatory property with respect to
$\phi$ or $p$. Differing from the usual classical pure L-design, the
energy of a mesoscopic quantum pure L-design can not be large than
$2\hbar / q_e^2 $.
Clearly, the lowest energy states are those states with
$ p = \phi + n h/q_e $,
then the eigenvalues of the electric current ( i.e.
$ \displaystyle { \frac{1}{L}\hat{P}  }$ ) 
of ground state can be obtained \cite{LiCh}.
The electric current on a mesoscopic  circuit of pure L-design
is not null in the presence of a magnetic flux (except
$\phi = nh/q_e$).
This is a pure quantum characteristic.
The persistent current in a mesoscopic L-design
is an observable quantity periodically depending on the flux $\phi$.
Because a mesoscopic metal ring is a natural pure L-design, the formula
we obtained is valid for persistent current in a single mesoscopic
ring \cite{Chand}. One can easily calculate the inductance of mesoscopic
metal ring, $L=8\pi r (\frac{1}{2}\ln\frac{8r}{a} - 1)$
where $r$ is the radius of the ring and $a$ is the radius of the metal
wire.
Then the formula for persistent current is
\begin{equation}
I(\phi) = \frac{\hbar}{8\pi r 
   (\frac{1}{2}\ln\frac{8r}{a} - 1)q_e}
     \sin(\frac{q_e}{\hbar}\phi).
\end{equation}
Differing from the conventional formulation of the persistent
current on the basis of quantum dynamics for electrons, our formulation
presented a  method from a new point of view. Formally, the $I(\phi)$
we obtained is a sine function with periodicity of
$\phi_0 =\displaystyle\frac{h}{q_e}$. 
But either the model that the electrons move freely in an ideal 
ring \cite{Cheung}, or the model that the electrons have hard-core
interactions between them \cite{LiMa2} can only give the sawtooth-type
periodicity. Obviously, the sawtooth-type function is only
the limit case for $q_e /\hbar \rightarrow 0 $.

\section{Quantum C-design and Coulomb Blockade}
In Coulomb blockade experiments,
the mesoscopic capacity may be relatively very small 
(about $10^{-8}F$) but the inductance of a macroscopic circuit 
connecting to a source is relatively large
because the inductance of a circuit is proportional to the
area which the circuit spans. We can neglect the term 
reversely proportional to $L$ in (\ref{eq:Hamiltonian}), 
and study the equation for a pure C-design.
\begin{equation}
\left(\frac{1}{2C}\hat{q}^2 -\varepsilon\hat{q}
  \right)|\psi>=E|\psi>.
\end{equation}
Apparently, the Hamiltonian operator commutes with 
the charge operator,
so they have simultaneous eigenstates. 
The energy for the eigenstate $|n>$ is 
$ E=(n q_e - C\varepsilon)^2/2C 
  -C\varepsilon^2/2$,
which involves both the charge quantum number and the voltage source.
After some analyses, we find the relation between charge $q$
and the voltage $\varepsilon$ for the ground state,
\begin{equation}
q=\sum_{m=0}^{\infty}
   \left\{ \theta[\varepsilon -(m+\frac{1}{2})\frac{q_e}{C}]
            -\theta[-\varepsilon -(m+\frac{1}{2})\frac{q_e}{C}]
             \right\}q_e
\label{eq:chargestep}
\end{equation}
where $\theta(x)$ is the step function. 
The corresponding eigenstate is 
\begin{equation}
|\psi(\varepsilon)>_{ground}=
   \sum_{-\infty}^{\infty}
     \left\{\theta[\varepsilon -(m-\frac{1}{2})\frac{q_e}{C}]
            -\theta[-\varepsilon -(m+\frac{1}{2})\frac{q_e}{C}]
             \right\}|m>.
\end{equation}
The dependence of the current on time is obtained by taking 
derivative 
\begin{equation}
\frac{dq}{dt}=\sum_{m=0}^{\infty}q_e
   \left\{\theta[\varepsilon -(m+\frac{1}{2})\frac{q_e}{C}]
            +\theta[\varepsilon + (m+\frac{1}{2})\frac{q_e}{C}]
             \right\}\frac{d\varepsilon}{dt}.
\end{equation}
Clearly, the currents are of the form of sharp pulses which occurs
periodically according to the changes of voltage. The voltage
difference between two pulses are $q_e/C$. This is the called Coulomb
blockade phenomena caused by the charge discreteness. 
If considering the problem in $p$-representation, we have
\begin{equation}
-\frac{\hbar^2}{2C}
  \left(\frac{d}{dp}-i\frac{C}{\hbar}\varepsilon
   \right)\tilde{\psi}(p)=
    (E + \frac{C}{2}\varepsilon^2 )\tilde{\psi}(p).
\end{equation}
Apparently, $\tilde{\psi}_{\beta}(p) \propto e^{i\beta p}$
solves the equation with the energy
$ E_{\beta}=\hbar^2 
  (\beta -\frac{C}{\hbar}\varepsilon)^2/2C
   -C\varepsilon^2/2 $,
from which we know the ground state has $\beta=C\varepsilon/\hbar$. 
So the wave function for the ground state is
$\tilde{\psi}_{ground}(p) = N e^{i\frac{C}{\hbar}\varepsilon p}$
($N$ is the normalization constant).

\bigskip

{\bf ACKNOWLEDGMENTS}

The author would like to thank Dr. B. Chen for the collaboration in
Zhejiang University; the supports by EPFL and by the Pao Yu-Kang \&
Pao Zhao-Long scholarship for Chinese scholars.

\end{document}